\documentclass[longabstract,preprint,dvips,12pt]{aastex}
\usepackage{natbib}
\usepackage{graphicx}
\usepackage[latin1]{inputenc}
\usepackage{amssymb}
\usepackage{xspace}
\usepackage{color}
\newcommand{\tsr}{{TSR}\xspace}
\newcommand{\nc}{N}
\newcommand{\eminus}{{\rm e}^{-}}
\newcommand{\kt}{k_{\rm B}T}
\newcommand{\ktpar}{\kt_{\parallel}}
\newcommand{\ktperp}{\kt_{\hspace*{-0.15em}{\perp}}}

\newcommand{\fetenplus}{\mbox{\ion{Fe}{11}}\xspace}

\newcommand{\feelevenplus}{\mbox{\ion{Fe}{12}}\xspace}
\newcommand{\feelevenplusc}{\mbox{Fe$^{11+}$}\xspace}

\newcommand{\e}[1]{\ensuremath{\times 10^{#1}}}			

\title{Electron-ion Recombination
	of \feelevenplus forming \fetenplus:
	Laboratory Measurements
 and Theoretical Calculations
}

\author{O. Novotn\'y\altaffilmark{1}, 
	N.~R.~Badnell\altaffilmark{2},
	D.~Bernhardt\altaffilmark{3},
	M.~Grieser\altaffilmark{4},
	M.~Hahn\altaffilmark{1}, 
	C.~Krantz\altaffilmark{4}, 
	M.~Lestinsky\altaffilmark{1,5},
	A.~M\"{u}ller\altaffilmark{3},  
	R.~Repnow\altaffilmark{4},
	S.~Schippers\altaffilmark{3},
	A.~Wolf\altaffilmark{4}, and
	D.~W.~Savin\altaffilmark{1}
	}

\date{\today}
\email{oldrich.novotny@mpi-hd.mpg.de}

\altaffiltext{1}{Columbia Astrophysics Laboratory, Columbia University, New York, NY 10027, USA}
\altaffiltext{2}{Department of Physics, University of Strathclyde, Glasgow G4 0NG, UK}
\altaffiltext{3}{Institut f\"ur Atom- und Molek\"ulphysik, Justus-Liebig-Universit\"at Giessen, D-35392 Giessen, Germany}
\altaffiltext{4}{Max Planck Institute for Nuclear Physics, 69117
Heidelberg, Germany}
\altaffiltext{5}{Present address: GSI Helmholtzzentrum f\"ur 
Schwerionenforschung mbH, D-64291 Darmstadt, Germany}


\begin{abstract}
We have measured electron-ion recombination for \feelevenplus forming \fetenplus
using a merged beams configuration at the heavy-ion storage ring \tsr located at
the Max Planck Institute for Nuclear Physics in Heidelberg, Germany. The
measured merged beams recombination rate coefficient (MBRRC) for collision
energies from $0$ to $1500$~eV is presented. This work uses a new method for
determining the absolute MBRRC based on a comparison of the ion beam decay rate
with and without the electron beam on. For energies below $75$~eV, the spectrum
is dominated by dielectronic recombination (DR) resonances associated with
$3s\rightarrow3p$ and $3p\rightarrow3d$ core excitations. At higher energies we
observe contributions from $3\rightarrow \nc^\prime$ and $2\rightarrow
\nc^\prime$ core excitations DR. We compare our experimental results to
state-of-the-art multi-configuration Breit-Pauli (MCBP) calculations and find
significant differences, both in resonance energies and strengths. We have
extracted the DR contributions from the measured MBRRC data and transformed them
into a plasma recombination rate coefficient (PRRC) for temperatures in the
range of $10^3$ to $10^7$~K.  We show that the previously recommended DR data
for \feelevenplus significantly underestimate the PRRC at temperatures relevant
for both photoionized plasmas (PPs) and collisionaly ionized plasmas (CPs). This
is to be contrasted with our MCBP PRRC results which agree with the experiment
to within 30\% at PP temperatures and even better at CP temperatures. We find
this agreement despite the disagreement shown by the detailed comparison between
our MCBP and experimental MBRRC results. Lastly, we present a simple
parameterized form of the experimentally derived PRRC for easy use in
astrophysical modelling codes.
\end{abstract}

\begin{document}
\doublespace
\maketitle

\section{Introduction}
\label{introduction}
Iron M-shell ions have been identified as the dominant source of the $15-17$
\AA\ absorption feature seen in \textit{Chandra} and \textit{XMM-Newton} X-ray
observations of warm absorbers in active galactic nuclei (AGNs; e.g.,
\citealt{Sako2001}, \citealt{Holczer2010}). This feature can be used to diagnose
the properties of AGNs \citep{Behar2001}. Such a study, however, requires
reliable low temperature dielectronic recombination (DR) rate coefficients for
iron M-shell ions, as has been discussed by \cite{Netzer2004},
\cite{Kraemer2004}, \cite{Chakravorty2008}, and \cite{Kallman2010}.  

A series of experimental and theoretical studies has been performed to meet
this need \citep{Gu2004, Badnell2006a, Badnell2006b, Altun2006, Altun2007,
Schmidt2006, Schmidt2008, Lukic2007, Lestinsky2009}. The measurements in this
series are based on a storage ring merged beams technique utilizing the \tsr
heavy ion storage ring located at the Max-Planck-Institute for Nuclear Physics
in Heidelberg, Germany \citep{Habs1989}. A bibliographic compilation of storage
ring DR measurements for astrophysically relevant ions has recently been given
by \cite{Schippers2009} and an overview of TSR experiments on Fe ions is given
by \cite{Schippers2010}.

As part of this effort, here we present new experimental results for P-like
\feelevenplus forming S-like \fetenplus.  Throughout the rest of this paper
recombining systems are identified by their initial charge state. The
most relevant channels for DR of \feelevenplus are
\begin{eqnarray}
\label{formel:FeXIIchannels}
{\rm Fe}^{11+}\ (3s^2\,3p^3\  [^4{S}^{\rm o}_{3/2}]) + \eminus
  \rightarrow \left\{ \begin{array}{ll}
    {\rm Fe}^{10+}\ (3s^2\,3p^3\, [{^2{D}}^{\rm o}_{3/2;5/2};\
{^2{P}}^{\rm o}_{1/2;3/2}]\, nl) \\
    {\rm Fe}^{10+}\ (3s\,3p^4\,nl) \\
    {\rm Fe}^{10+}\ (3s^2\,3p^2\,3d\,nl)\\
    {\rm Fe}^{10+}\ (3s\,3p^3\,3d\,nl)
    .
  \end{array} \right.
\end{eqnarray}
The incident electron is captured into a Rydberg level with a principal
quantum number denoted by $n$. DR proceeds via excitation of a core electron
with a principal quantum number which we denote by $\nc$.
The energies of the core excitations
corresponding to $\Delta \nc = \nc'-\nc =0$ DR are listed in Table
\ref{table:fe11energylist}. 
\feelevenplus is predicted to form at plasma temperatures of $\log T_{\rm
e}({\rm K}) \sim 4.76-5.48$ in photoionized gas \citep{Kallman2010} and $\sim
5.87-6.25$ in collisionally ionized gas \citep{Bryans2006,Bryans:ApJ:2009} where
$T_{\rm e}$ is the electron temperature in Kelvin.

The remainder of this paper is organized as follows:
Section~\ref{theory} gives a brief summary of the theoretical
calculations.  Section~\ref{exp_setup} describes the experimental
setup used here.  Our experimental results for the merged beams
recombination rate coefficient (MBRRC) are presented and compared to
theory in Section \ref{results_mbrrc}.  Section~\ref{results_plasma}
reports our experimentally-derived DR plasma recombination rate
coefficient (PRRC), a comparison with theory, and a simple fitting  formula for
plasma modeling.
Lastly, a summary is given in Section~\ref{conclusion}.

\section{Theory}
\label{theory}

The partial, energy averaged, DR cross section 
$\bar{\sigma}^{z}_{fi}$ from an initial state $i$ of an ion $X^{+z}$
into a resolved final state $f$ of an ion $X^{+z-1}$  is given in the isolated
resonance approximation by \citep{BBGP1, Badnell2006a}
\begin{equation}
\bar{\sigma}^{z}_{fi}(E_{\rm c})={2\tau_{0}(\pi a_0 I_{\rm H})^{2} \over
E_{\rm c}\, \Delta
E}
\sum_j{\omega_{j} 
\over \omega_{i}}\,
 { \sum_{l}A^{\rm a}_{j \rightarrow i, E_{\rm c}\,l} \, A^{\rm r}_{j
\rightarrow
f}
\,\over \sum_{h} A^{\rm r}_{j \rightarrow h} + \sum_{m,l} A^{\rm a}_{j
\rightarrow m, E_{\rm c}\,l}}.
\label{eqPDR}
\end{equation}
Here $\omega_j$ is the statistical weight of the doubly-excited resonance state
$j$ in the recombined $+z-1$ ion, $\omega_i$ is the statistical weight of
the initially state of the initial $+z$ ion, and the
autoionization ($A^{\rm a}$) and radiative ($A^{\rm r}$) rates are in inverse
seconds. The indices $h$ and $m$ are for states in the $+z-1$ and $+z$ ions,
respectively. $E_{\rm c}$ is the energy of the incoming continuum electron (with
orbital angular momentum $l$) which is fixed by the position of the resonances,
$\Delta E$ is an arbitrary bin width, $I_{\rm H}$ is the ionization potential
energy of the hydrogen atom, $\tau_0$ is the atomic unit of time, and
$a_0$ is the Bohr radius.

We use the general atomic collision code  {\sc autostructure}~\citep{AS,
Badnell2006a, DW} to calculate the constituents of Equation \ref{eqPDR}. The
calculations for  $\Delta \nc =0$ core-excitations were carried out in
intermediate coupling using a configuration interaction expansion for the
Fe$^{11+}$ 15-electron target involving a Ne-like core and valence
configurations of $3s^2\,3p^3$, $3s\,3p^4$, $3s^2\,3p^2\,3d$, $3p^5$,
$3s3p^3\,3d$, $3s^2\,3p3d^2$, $3p^4\,3d$, and $3s3p^2\,3d^2$ to which continuum
and Rydberg electron orbitals were coupled. The 16-electron configurations,
formed by adding a $3s$, $3p$, or $3d$ orbital to the 15-electron
configurations, were included to describe outer electron radiative transitions
into the core. Radiative transitions from higher $n$ levels were described
hydrogenically. The merged beams experiment does not resolve the final state and
so all of the results that we present are for the total recombination cross
section, i.e., summed over all $f$ that are stable against autoionization and
are not field ionized in TSR before they are detected. These energy-averaged
cross sections can be convolved with the experimental energy distribution for
comparison with the measurements. They can also be convolved with a Maxwellian
distribution for modelling use and summed over all possible stable final states
to generate a total PRRC. For $\Delta N = 0$ DR, the sum over the Rydberg $nl$
states extended to $n=1000$ and $l=11$ for the total Maxwellian rate
coefficients while for comparison with experiment the relevant survival
probabilities were folded into the sum over the final-states (e.g.,
\citealt{Schippers2001}).

The contributions from $\Delta \nc >0$ core-excitations were also calculated by
{\sc autostructure} but using a configuration-averaged
approximation~\citep{CA}. This approximation is only suited for $\Delta \nc >0$
Maxwellian rate coefficients since it only resolves resonance positions and
channels by configuration only. The omission of configuration mixing is
not a severe
one for the total PRRC given that mixing conserves the overall amount
of resonance strengths  and, at the energies relevant here, causes only 
small fractional errors in the resonance energies~\citep{Sn}. We include
both $\nc=2\rightarrow 3$ and $\nc=3\rightarrow 4$ core excitations.  The sum
over the captured electron $nl$ Rydberg states extended to $n=100$ and $l=6$ for
these total Maxwellian rate coefficients. No difference is seen between the
calculations with and without the field ionization effects included.

\section{Experimental Setup}
\label{exp_setup}

\subsection{General}
\label{lExpGen}

Measurements were performed using the heavy-ion storage ring \tsr.
Details on the various aspects of the merged beam technique as used at
TSR have been described at length by \cite{Kilgus1992},
\cite{Lampert1996}, \cite{Pastuszka1996}, \cite{Schippers2001},
\cite{Wolf2006}, \cite{Lestinsky2008}, \cite{Schmidt2008} and
\cite{Lestinsky2009}.  Here we discuss only those aspects particular
to the present work.

A 150~MeV beam of $^{56}$\feelevenplusc was generated by first passing
$^{56}$Fe$^-$ ions through a carbon foil to strip and produce the
desired charge state and then further accelerating them.  After
charge-to-mass selection, the Fe$^{11+}$ beam was injected into the storage
ring.  Ions were accumulated by multi-turn injection and ``e-cool
stacking'' \citep{Grieser1991}.  Typical stored ion currents were
$\sim 1-2~\mu$A during data acquisition with storage times of 
$\approx 10$~s.

Ions produced by foil-stripping can be highly excited 
\citep{MartinsonGaupp1974}.  Here we stored the ions for $\sim
1.5-2.0$~s before beginning data acquisition.  We used a numerical
model of the radiative decay process to determine the level
populations in the ion beam after this initial storage time.  The
model considered excited states up to the $3s^2\,3p^2\,4d\ ^2S_{1/2}$ level and
included all 66 transitions, involving 31 levels, for which radiative
rates are given in the ASD/NIST database \citep{AtomicSpectraDatabase}.
These data, in turn, come from \cite{fawcett_classification_1972},
\cite{huang_energy-level_1984}, and \cite{shirai_spectral_1990}.  The
initial relative populations of the excited states were modelled using
a Boltzmann distribution with a temperature of $k_{\rm B}T=750$~eV.  This
corresponds to the approximate collision energy of the foil electrons
as the Fe$^-$ ions passed through the carbon foil.  After 1.5 s of
storage, over 98.5\% of the ion beam is expected to be in the ground
state.  A factor of 10 increase or decrease in the effective
temperature has an insignificant effect on this estimate.  The most
critical lifetime determining the final population is that of the
$3s^2\,3p^3\ [^2{D}^o_{5/2}] \rightarrow 3s^2\,3p^3\ [^4{S}^o_{3/2}]$
radiative transition to the ground state.  The Einstein coefficient
listed in ASD/NIST database at 1.84~s$^{-1}$ is smaller compared to
the more recent experimental value of 3.26~s$^{-1}$ reported by
\cite{trabert2002}. Using the latter result, the predicted ground state
population after 1.5~s of storage is 99.0\%. For both lifetimes,
the ground state population averaged over the entire $\sim 20$~s storage time is
greater than $99.9$\%. 

TSR is equipped with two different electron beam devices located in separate
sections of the ring.  Each electron beam can be merged to co-propagate with the
stored ions.  One of the devices is called the Cooler \citep{Steck1990} and the
other the Target \citep{Sprenger2004}. Either or both of the electron beams can
serve to reduce the energy spread of the ions, i.e., to cool the ions.  Electron
cooling \citep{Poth1990}
results in a narrow ion beam diameter ($<1$~mm) with a low energy spread.
Additionally, either one of the electron beams can be used as an interaction
medium while the other continues to cool the ion beam.  Electron-ion collisions
can then be
investigated by varying the energy of one of the electron
beams.

The electron beam energy spread is described by a flattened Maxwellian
distribution characterized by the longitudinal and transverse
temperatures $T_{||}$ and $T_\perp$ \citep{Kilgus1992}.  At a
collision energy of $\hat{E}$, the corresponding center-of-mass
collision energy resolution $\Delta \hat{E}$ is approximately given by
$\Delta \hat{E} = [(\ln(2)\ktperp)^2 + 16\ln(2)\hat{E}\ktpar]^{1/2}$
\citep{Mueller1999}.  The Cooler uses a thermionic emission cathode.  Typical
electron beam temperatures are $\ktpar^{\rm{c}}\approx180~\mu$eV and
$\ktperp^{\rm{c}}\approx13.5$~meV \citep{Lestinsky2008}.  The Target uses a
photocathode \citep{Pastuszka2000,Orlov2004a}.  From this we produce a beam with
significantly lower temperatures of $\ktpar^{\rm{t}}\approx25~\mu$eV and
$\ktperp^{\rm{t}}\lesssim1.5$~meV \citep{Lestinsky2008}. The complexity of the
\feelevenplusc DR spectrum prevented direct determination of the Cooler and
Target temperatures from the measured spectrum.  Hence, for the results
presented here we used values from a similar experiment
\citep{Lestinsky2008}.

The products of charge-changing reactions are deflected from the
parent ion beam by the first dipole magnet downstream of each electron
beam device and are directed onto a detector.  Scintillator detectors
for measuring recombination are located after both the Cooler and the
Target \citep{Miersch1996,Wissler2002,Lestinsky2007}.  To measure
electron impact ionization (EII), we used a converter plate coupled
with a channel electron multiplier
\citep[CEM;][]{rinn_1982,Linkemann1995} located after the Cooler.  The
recombination and ionization signals were used to determine the
absolute recombination rate coefficient from the ion beam lifetime as
described below.

The efficiency of each detector, in the absence of dead time effects,
is essentially 100\%.  The dead time for each detector was estimated
from the maximal width of the electronic pulses.  This was shorter
than 100~ns in all cases.  As the count rates never exceeded 300~kHz,
the corresponding dead time corrected detector efficiency did not
decrease below 97\%.

Field ionization of the recombined ions in the dipole magnet can ionize
electrons captured into Rydberg levels  with $n\ge n_{\rm cut}$. A
semi-classical calculation yields $n_{\rm cut}=42$.  However, during the travel
time from the interaction region to the dipole, some of the initially high
Rydberg states can radiatively decay below $n_{\rm cut}$ and avoid 
ionization in the magnet. \cite{Schippers2001} have derived a model to
calculate $nl$-specific detection probabilities taking into account the field
ionization and radiative de-excitation processes.  The average Rydberg state
cut-off resulting from this model is $n_{\rm cut}\approx54$. For comparison of
our results to the theory we use the detection probabilities provided
by this model.

\subsection{Determination of the relative MBRRC spectra}
\label{l:relative}

Normally the absolute MBRRC is derived from the measured recombination counts
using an appropriate normalization to the electron density and ion current
\citep[e.g.,][]{Schmidt2008}.  The electron density can be readily measured
accurately \citep[e.g.,][]{Lestinsky2009}. However, here the average
stored ion current of $\sim 1-2~\mu$A in the present experiment was too low to
be directly measured using the DC current transformer installed in the ring.
Instead, a relative MBRRC was determined by normalizing the
signal count rate to a proxy for the ion current.  For this we used the relative
intensity of the ion beam as recorded by a beam profile monitor
\citep{Hochadel1994}.  To derive the absolute calibration of the MBRRC spectrum
we use the approach described in Section~\ref{l:abs}.

Data were collected for electron beam laboratory energies from $\approx 1450 -
6000$~eV.  These translate to center-of-mass collision energies $0 \le
\hat{E} \le 1500$~eV, where $\hat{E}=0$ corresponds to matched electron
and ion beam velocities.   The full range was covered using
the Cooler to collect data with continuous cooling by the Target.  The
high electron beam density of $n_{\rm e} \approx 2.7
\times 10^7$~cm$^{-3}$ in the Cooler allowed for shorter data acquisition times
than
that of the Target which had a density of $n_{\rm e} \approx 1.1 \times
10^6$~cm$^{-3}$.  The roles were reversed to cover the low energy
range $0 \le \hat{E} \le 0.2$~eV at the higher energy resolution
offered by the Target.  As discussed in Section~\ref{results_mbrrc},
these data aided in the extraction of the low energy DR resonance
strengths needed to generate a reliable low temperature PRRC.

Each data run lasted $\sim 1-2$~hrs, during which time we continuously repeated
the measurement cycle of ion injection, cooling, and data acquisition.  During
data collection the electron beam energy was stepped between a variable
\emph{measurement} energy and a fixed \emph{reference} energy, creating $\sim
350$ measurement pairs of typically $\lesssim 20$~s total duration.  The
measurement energy was changed after each reference
step. For each data run the measurement energy range in the laboratory
frame spanned over $\sim 140 - 1000$~eV.  This is much smaller than the total
laboratory energy range studied. The reason for splitting to
smaller energy ranges is related to the required settling time of the power
supplies when switching the electron beam laboratory energy to measurement or
reference. In order to keep this time short (15~ms in our case), the reference
laboratory energy in each run was kept close to the measurement laboratory
energy range. As a result, several data runs were needed to cover the entire
laboratory energy range measured.  After the 15~ms settling time, the subsequent
dwell time at each measurement or reference energy step was $10-25$~ms.  

The signal at reference consists of background due to electron capture from
residual gas, radiative recombination (RR), and potentially also DR. At high
energies DR is negligible and so when we subtracted the reference from the
measurement it was only necessary to re-add the small theoretical RR
contribution, thereby insuring that only the appropriate background was
subtracted. In the low energy runs, however, the DR contribution at the
reference energy became non-negligible. We corrected for this by comparing a
lower energy run to an overlapping higher energy run and shifting the offset in
the former to match the data in the latter.

\subsection{Absolute scaling of MBRRC spectra}
\label{l:abs}

We put our relative MBRRC results on an absolute scale
using measurements of the ion beam lifetime with the Cooler electron beam first
\emph{off} and then \emph{on}.  A similar normalization method has been used
for molecular ion studies \citep{Pedersen2005}.  Our approach builds upon and
extends their work.

With the Cooler off, the ion beam decays exponentially due to
collisions with the residual gas along the entire TSR circumference.
The measured storage lifetime
$\tau^{\rm off}$ is inversely proportional to the loss rate $\lambda_{\rm
  res}^{\rm off}$ as
\begin{equation}
\frac{1}{\tau^\mathrm{off}} = \lambda_\mathrm{res}^\mathrm{off}.
\label{eqoff}
\end{equation} 
With the Cooler on at a fixed energy $\hat{E}$, the measured storage
lifetime is now due to both collisions with the residual gas and also to
RR and DR in the Cooler giving
\begin{equation}
\frac{1}{\tau^\mathrm{on}} =  \lambda_\mathrm{res}^\mathrm{on} 
+  \hat{\alpha}\,n_e\,\eta.
\label{eqon}
\end{equation} 
Here $\lambda_{\rm res}^{\rm on}$ is the residual gas collisional loss rate
with the Cooler on, $\hat{\alpha}$ is the RR+DR MBRRC at 
$\hat{E}$, $n_e$ is the Cooler electron density,
and $\eta=L/C$ is the overlap length $L$ of the ion and electron beams
normalized by the total storage ring circumference $C$.  As a first
approximation, we set $\lambda_\mathrm{res}^\mathrm{off} =
\lambda_\mathrm{res}^\mathrm{on}$ and solve Equations~\ref{eqoff} and
\ref{eqon} to yield an absolute MBRRC
\begin{equation}
\hat{\alpha}(\hat{E}) = 
\frac{1}{n_e\,\eta}
\left(\frac{1}{\tau^\mathrm{on}}-\frac{1}{\tau^\mathrm{off}}\right).
\label{eqabs1}
\end{equation} 
A more thorough derivation, accounting for the slight differences
between $\lambda_\mathrm{res}^\mathrm{off}$ and
$\lambda_\mathrm{res}^\mathrm{on}$ due to changes in the pressure of residual
gas, is given in Appendix~\ref{l:lifetime}.
Here, this difference results in less than a 5\% change in
$\hat{\alpha}(\hat{E})$.  The relative MBRRC results of
Section~\ref{l:relative} can then be put on an absolute scale
by adjusting the data so that the value at $\hat{E}$ matches the absolute rate
derived using this lifetime method.

Figure~\ref{lfAbs} shows an example of data collected using this
method for matched ion and electron beam velocities
($\hat{E}=0$~eV). For these results, the ions were cooled
for 3~s after injection and then the Cooler beam switched off. Some
time later the Cooler beam was switched back on.  The relative beam
intensity for each phase was monitored by detecting products from
ion collisions resulting in electron capture and ionization.
We determined the beam lifetime using the decaying signal
strength on both the recombination and ionization detectors.  For the
results presented here the Target electron beam was on
continuously.

For the present work the measured beam lifetime decreased dramatically
when the Cooler was turned on due to the extraordinary high rate
coefficient of DR+RR at $\hat{E}=0$ which dominates over
collisions with the
residual gas at the given electron density.  This can be seen in both
the recombination and
ionization detector count rates shown in Figure~\ref{lfAbs}.  The Cooler
energy is below the \feelevenplus threshold for electron impact ionization
($\sim330$~eV, \citealt{AtomicSpectraDatabase}) and so the signal
on the ionization detector originated exclusively from electron stripping in
ion collisions with residual gas. Thus,  we attribute the small increase
seen in the ionization product count
rate when the Cooler is turned on
as being due to a corresponding increase in the residual gas pressure.  This
pressure change is accounted for in our analysis as is
described in Appendix~\ref{l:lifetime} and has less than a 5\% effect on our
results.  We also found that, to within the statistical errors, the absolute
scaling method used here gave the same results independent of when the Cooler
was switched on or whether the Target was on or off.

We use the absolute MBRRC results to scale our relative MBRRC data.  
These are then corrected for the effects due to the merging and
demerging of the electrons with the ions \citep{Lampert1996}.  This
correction largely removes errors due to the uncertainty in the
exact electron-ion beams overlap length.

\subsection{Uncertainties}

The $1\sigma$ statistical error in our rate coefficient data is about 1\% for
collision energies below 1~meV.  As the energy increases, the recombination
rate, and hence the signal rate, decreases.  This leads to an increase in the
statistical error with increasing energy.  It remains below 5\% for energies
up to 66~eV and $\sim 7\%$ for $66-1500$~eV.

We treat the various systematic errors in our measurement as
uncorrelated and add them in quadrature.  The resulting $1\sigma$
systematic error is estimated to be 12\%, 13\%, and 40\% for collision
energies of 0~eV, 66~eV, and above 66~eV, respectively.  Here we
briefly review the sources of the total systematic uncertainty.  Further
details about systematic errors can be found in the TSR references
cited in Sec.~\ref{lExpGen}.

The largest source of systematic error below 66~eV is due to the electron
density determination.  The unusually high 10\% error for the data
here resulted from the accidental use of a degraded
photocathode for the absolute MBRRC measurement.  The reproducibility
of the absolute results and the extraction of ion beam lifetimes
for the  determination of the absolute MBRRC at 0~eV result in an
additional 5\% uncertainty.

The data have been stitched together going from high to low energies to correct
for the changing reference energy.  As the data have been normalized at
$\hat{E}=0$, this stitching results in a 5\% error at 66~eV and up to a
35\% error at higher energies. The large increase in this error above 66~eV is
due to the $\sim100$ times decrease in the magnitude of the MBRRC. Other
remaining sources of error include the corrections for the merging and demerging
of the beams \citep[$1\%$;][]{Lampert1996} and the deadtime counting
efficiencies of detectors.

\section{Merged beams recombination rate coefficient}
\label{results_mbrrc}

The measured MBRRC data are displayed in Figures~\ref{lfMBRC0} and \ref{lfMBRC1}
for the energy ranges $0-80$~eV and $65-1500$~eV, respectively.  These data were
acquired using the Cooler as the probe beam and the Target for cooling.  Also
shown are our {\sc autostructure} MCBP results, with and without the
experimental field ionization effects.  The theoretical cross section has
been multiplied by the collision velocity and convolved with the Cooler energy
spread to generate a theoretical MBRRC. 

The \feelevenplus resonance spectrum for this system with a half open
$p$-shell is very rich and challenging to
disentangle.  In general the features are broad and unresolved making
individual assignments essentially impossible.  DR via $3s^23p^3$
intra-configuration core excitations is expected for energies below
$\approx 10$~eV.  The bulk of these contribute significantly only below $\approx
5$~eV as can be seen by the step-like drop in the MBRRC at this energy.  Moving
up in energy, most of the $3s 3p^4$ core excitations are expected to occur
below $\approx 35$~eV.  At energies of $35-75$~eV, the features become
more regular. These are due largely to $3s^2 3p^2 3d$ core excitations
and the resonances can be more easily assigned.  For clarity we have labeled
only those resonances which are due to the strongest $3s 3p^4$ and $3s^2 3p^2
3d$ series.  Filling in the many other resonances series would make the
figures too cluttered for meaningful inspection.

Given the complexity of the spectra, for comparison with theory
we have followed \cite{Lestinsky2009} and calculated
\begin{equation}
\kappa = 
{\int\alpha^{\rm DR}_{\rm{theo}}dE
\over
\int \alpha^{\rm DR}_{\rm{exp}}dE}
\label{eq:kappa}
\end{equation}
for sequential energy ranges.  The lowest energy considered is 13.5~meV to avoid
the well known effects of enhanced RR near $\hat{E}=0$
\citep{Gwinner2000,Wolf2003,Hoerndl2006}.  In the denominator of Equation
\ref{eq:kappa} we take $\alpha^{\rm DR}_{\rm exp} = \alpha_{\rm exp} -
\alpha^{\rm RR}_{\rm theo}$.  We have calculated $\alpha^{\rm RR}_{\rm theo}$
using both the hydrogenic Bethe-Salpeter method \citep{Hoffknecht2001} and a
hydrogenic quantum mechanical dipole approximation for low $n$ and a
semiclassical approach with Stobbe corrections for high $n$
\citep{Stobbe:AP:1930}.  The difference in $\alpha^{\rm RR}_{\rm theo}$ between
the two methods is insignificant.  

We find mixed agreement between theory and experiment. Results for $\kappa$ are
given in Table~\ref{table:fe11integrals}.  If the difference were solely due to
the estimated $1\sigma$ experimental systematic error, we would expect
this ratio to range
between $0.88-1.15$ and $0.71-1.67$ for collision energies below and above
66~eV, respectively.  In the energy range $0.0135-0.45$~eV, $\kappa$ is nearly 
one third.  This is most likely due to incorrectly
predicted
resonance energies resulting from the well-known difficulty of calculating DR
resonance positions at low energies (cf., \citealt{Schippers2009} and
\citealt{Schippers2010} and references therein). In the various energy ranges
between 0.45~eV and 46.0~eV, theory is smaller by more than the $1\sigma$
experimental systematic uncertainty.  Reasonable agreement is found in the range
$46-53$~eV.  However in the range $53-59$~eV, theory is 1.4~times greater than
experiment.  This apparent systematic overestimate of the integrated theoretical
resonance strength occurs for $\Delta \nc=0$ DR where the radiative
stabilization is primarily by the core electron and the Rydberg electron
occupies $n \gtrsim 10$. Similar discrepancies have been seen in previous work
\citep{Lestinsky2009} and are discussed in more detail by \cite{Lestinsky2012}. 
The range $59-66$~eV includes six Rydberg series, mostly with a $3s^{2}3p^{2}3d$
configuration, and the large $\kappa$ might be partly explained by uncertainties
in the model for the experimental field ionization effect. Lastly, the range
$66-75$~eV covers ten Rydberg series limits associated with the $3s^{2}3p^{2}3d$
configuration. The cause of the low $\kappa$ here is unclear.

The MBRRC at energies above 75~eV is dominated by $\Delta \nc>0$ DR. Here
the MBRRC is $\sim 100$~times weaker than for $\Delta \nc=0$ at lower
energies. Not surprisingly, our configuration averaged calculations do a poor
job of reproducing the observed resonance structure. Between $\sim 75 - 330$~eV
the spectrum is expected to be dominated by DR via $3 \to \nc^\prime$ core
excitations where $\nc^\prime \ge 4$.  The resonances between $\sim
75-217$~eV we attribute to $3\rightarrow 4$ excitations and the region between
$\sim 217$~eV and the ionization limit of excitations into $\nc'\geq5$. In
this first range, we find $\kappa = 0.68$. We cannot determine $\kappa$ for the
$217 - 330$~eV range  as the $\nc^\prime \geq 5$ resonances were not
included in the theoretical model. The small decrease in the DR signal at $\sim
330$~eV corresponds to the $\nc=3$ ionization threshold
\citep{AtomicSpectraDatabase}. The next range from $\sim
330 - 855$~eV is dominated by DR via $2 \to \nc$ core excitations where
$\nc^\prime \ge 3$. We attribute the resonances seen in this range mainly to
$2\rightarrow 3$ excitations and calculate a $\kappa$ of 0.28. For the range
$\sim 855 - 1073$~eV, no theoretical data exists and we are unable to determine
$\kappa$.
The $2\rightarrow \nc'$ channels cease to
contribute to DR once ionization from the $\nc=2$ level becomes possible at
1073~eV \citep{Kaastra:AAS:1993}, as is readily visible in the measured data. 
It is worth noting, too, that a significant amount of the measured DR flux above
$\sim 60$~eV is not  accounted for in the theoretical calculations.

We have also measured the MBRRC using the Target as the probe beam and the
Cooler for cooling.  These results are shown in Figure~\ref{lfMBRC3} along with
the data collected using the Cooler. The differences seen below 0.001~eV are
attributed to enhancements in the RR signal as we discuss in Section 5.  At
higher energies, the greater resolution of the Target compared to the Cooler
allows additional resonance features to be resolved, particularly for collision
energies between 0.001~eV and 0.02~eV. Note also how the Target data drop to the
level of the RR background around 0.03~eV.  In Section~\ref{results_plasma}
these Target results are used to determine the contribution of the low energy DR
resonances to the PRRC.

\section{Plasma DR rate coefficients}
\label{results_plasma}


The derivation of the PRRC from the experimental MBRRC data has been discussed
in detail in \cite{Schippers2001, Schippers2004}, \cite{Schmidt2008}, and
\cite{Lestinsky2009}.  Four points need to be considered.  First, at
sufficiently high collision energies, the required cross section can be
extracted by dividing the MBRRC data by the classical relative velocity
$v_{\rm r} = \sqrt{2\hat{E}/m_{\rm e}}$.  For this we used the Cooler data shown
in Figure~\ref{lfMBRC0}.

Second, we need to account for recombination near various series
limits into those levels which are expected to be field ionized in our
experimental arrangement.  Here we took the difference between the
{\sc autostructure} calculations with and without field ionization effects,
scaled the difference by the $\kappa$ factor for the energy range just
below that where field ionization is an issue, and added the results
to our measured MBRRC.  The cross section was then extracted as described
above.

Third, at lower energies (here $\lesssim 0.11$~eV), the experimental energy
spread becomes comparable to $\hat{E}$.  Our approximation for the cross section
breaks down and one must fit the data to extract resonance energies and
strengths for the many unresolved resonances in the data. Here we used
data collected with the Target as shown in
Figure~\ref{lfMBRC3}.  The Target provides much higher resolution data for
extracting resonances strengths compared to the Cooler data. The fitting
procedure was described in detail by \cite{Schippers2004}, \cite{Schmidt2008},
and \cite{Lestinsky2009}.

Lastly, at near zero energy, the RR signal may be experimentally
enhanced \citep{Gwinner2000,Wolf2003,Hoerndl2006}. There may also be
unresolved DR resonances. Previous work has shown that experimental
enhancement of the RR MBRRC amounts to a factor of $\sim1.5 - 3$ as compared to
that predicted by RR theory. Any remaining difference is attributed to
unresolved DR resonances. As shown in Figure~\ref{lfMBRC3}, the differences seen
here at $\sim 10^{-4}$~eV are a factor of $\sim 360$ for the Cooler and $\sim
420$ for the target.  
These factors strongly suggest the presence of unresolved low energy DR
resonances. 
Also the fact that the slopes of the Target and Cooler DR data at $\sim
10^{-3}$~eV differ significantly from those predicted by RR theory supports this
hypothesis. 
We account for this likely
DR when fitting our MBRRC data by including resonances at energies of 0.08 and
0.7~meV.  We calculated the PRRC with and without these resonances.  The average
of the two PRRC results was used and half the difference
between the two taken as the uncertainty.  In this way we estimate the
uncertainty due to the unclear origin of these resonances. This error was then
propagated quadratically into the total error budget of the PRRC. However, the
contribution of these resonances to the total PRRC is insignificant above
$10^3$~K which includes the temperatures where \feelevenplus is predicted to
form in either photoionized or collisionally ionized gas (see
Figure~\ref{fig:fe11_plasma}).

Taking all these points into account, we have derived the PRRC following the
procedure laid out in \cite{Schippers2001,Schippers2004}, \cite{Schmidt2008},
and \cite{Lestinsky2009}. Figure~\ref{fig:fe11_plasma} shows the results for
\feelevenplus forming \fetenplus in the temperature range of $10^3-10^7$~K.  
The total uncertainty at an estimated $1\sigma$ level
reaches $\lesssim15$\% at $10^3$~K, $\sim15$\% at $10^5$~K, $\sim20$\% at
$10^6$~K, $\sim 27\%$ at $2 \times 10^6$~K, and $\sim45$\% at $10^7$~K.
Over the temperature range shown in the figure, the
experimental DR PRRC is $\gtrsim 35$ times larger than the theoretical RR value
of \cite{Badnell2006b}.

The temperature ranges where the fractional abundance of \feelevenplus is $\ge
1\%$ of the total Fe abundance in photoionized plasmas (PPs) and in
collisionally ionized plasmas (CPs) are indicated in
Figure~\ref{fig:fe11_plasma} as grey shaded areas \citep{Bryans2006,
Bryans:ApJ:2009,Kallman2010}.  Also plotted is the previously recommended DR
rate coefficient of \cite{ArnaudRaymond1992}.  These data significantly
underestimate the DR PRRC at temperatures of below $2 \times 10^5$~K, which are
of particular importance for PPs.  Also at temperatures relevant to CPs, their
recommended DR data are up to about $2.4$ times lower than our experimental
results. Similar behavior has also been seen for other M-shell iron ions
\citep{Schmidt2006, Lukic2007, Schmidt2008, Lestinsky2009}.  The data reported
by \cite{ArnaudRaymond1992} represent a compilation of theoretical calculations
largely from the 1970s and 1980s.  We attribute the differences seen, in part,
to the limitation of computer power at that time and the required approximations
necessary to make the calculations tractable.  More recent, state-of-the-art
calculations have been performed by \citet{Badnell2006b}. However, these do not
include DR via $\Delta N > 0$ core excitations.  We have extended those results
by including DR via $2 \to 3$ and $3 \to 4$ core excitations. At PP temperatures
both calculations are in significantly better agreement with experiment but
still differ by up to 30\% which is outside of the $1\sigma$ experimental error
bars.  At CP temperatures both sets of theoretical results agree with experiment
to within the experimental uncertainty, despite the significant disagreement
between the MCBP theory and our experimental data on the MBRRC level (Figure
\ref{lfMBRC0}). Obviously the averaging over the Maxwellian temperature
distribution leads to a washing out of the discrepancies on the MBRRC level.
Both the theoretical and experimental results indicate that $\Delta \nc>0$
channels contribute $\ge10$\% to the PRRC at CP temperatures and up to 20\% at
$10^7$~K.

To facilitate the use of our experimentally-derived PRRC in plasma
models, we have parameterized the data using the function
\begin{equation}
  \label{eqn:plasmafit_DR}
  \alpha^{\mathrm{fit}}_{\mathrm{P}}\,(T) = T^{-3/2}\, \sum\limits_i c_i
  \exp(-E_i/T).
\end{equation}
The fitted parameters $c_i$ and $E_i$ are listed in Table
\ref{table:plasmafit}.  The fit accurately reproduces the experimentally derived
PPRC to better than 2\% over the temperature range of $10^3-10^7$~K.  

\section{Summary}
\label{conclusion}

We have measured the MBRRC for DR of \feelevenplus forming \fetenplus
over the collision energy range of $0-1500$~eV.  A merged electron-ion beams
configuration was used at the TSR heavy ion storage ring.  Poor
agreement is found between the experimental and theoretical resonance
structure, particularly for energies below $\sim 35$~eV.  Significant
differences are also found for the integrated resonance strengths over
most of the measured energy range.  Similar discrepancies between
experiment and theory have been seen in our previous studies of DR for
Fe M-shell ions \citep{Schmidt2006, Lukic2007, Schmidt2008,
  Lestinsky2009}.

From our experimental results we have derived a DR PRRC for plasma
temperatures of $10^3-10^7$~K.  This range includes the temperatures
where \feelevenplus is predicted to be abundant in photoionized
and collisionally ionized cosmic plasmas, respectively.  In general we
see behavior
similar to that noted for DR of other Fe M-shell ions
\citep{Schmidt2006, Lukic2007, Schmidt2008, Lestinsky2009}.  The
previously recommended DR data of \cite{ArnaudRaymond1992}
underestimate the DR PRRC by orders of magnitude at temperatures
relevant for photoionized plasmas and by up to a factor of 2.4 for
collisionally ionized gas.  Much better agreement is found with
state-of-the-art MCBP theory, though significant differences do remain
at the lowest temperatures where modern theory is known to still have difficulties
accurately predicting the energies of the relevant DR resonances.

\acknowledgments
We thank the accelerator and \tsr group for their excellent support
during the beam time.  M.L., M.H., O.N.\ and D.W.S.\ were supported in
part by the NASA Astrophysics Research and Analysis program and the
NASA Solar and Heliospheric Physics Supporting Research program.

\appendix
\section{Lifetime based method for absolute scaling of MBRRC}
\label{l:lifetime}

As discussed in Sec.~\ref{l:abs}, measurements for the lifetime of the
stored ions with the Cooler \emph{off} and \emph{on} can be used to derive an
absolute MBRRC.  
However, as a consequence of desorption from the surface of the vacuum
chamber in the collector section, the residual gas pressure in the Cooler
increases when the electron beam is turned on.
Monitoring the pressure inside TSR by pressure gauges is not precise and local
enough to describe such increases. Here we explain how to account for these
pressure changes without relying on direct pressure measurements.

Our measurements are performed on ground state Fe$^{11+}$ at collision energies
below the threshold for electron-impact ionization. Essentially the only
electron-driven, charge-changing reaction which can occur under these conditions
is electron-ion recombination.  We estimate as insignificant the contributions
from electron impact excitation to a bound level followed by an ionizing
collision on the residual gas in the ring.  The ion beam is also free of
metastables, i.e., the beam composition does not change during measurements. 
The only other significant processes affecting the lifetime of the stored ions
are collisions with residual gas particles in the ring leading to electron
capture (recombination) or loss (ionization).

The Cooler energy is kept constant during the on-phase to insure that
the electron collision rate coefficient to be derived is constant
during measurement.  Detectors downstream of the Cooler are used to
monitor the various collision end-products.  The other electron beam
device in the ring needs to be continuously on or continuously off
so as not to disturb the measurement.  The derivation presented here assumes
that the Target is on continuously.  Also, although we collected data using the
Cooler, the role of the Cooler and Target may be readily interchanged.

The rate coefficient for the electron-ion recombination is
given by $\alpha^1$.  The rate coefficients for this reaction in the Cooler
(c) and Target (t) are generally not identical, $\alpha^1_{\rm c} \ne
\alpha^1_{\rm t}$, as each device has a different electron beam energy
spread.  Additionally, each device can be operated at different
energies.

Collisions of the stored ions with the residual gas in the ring can result
in either electron capture from the gas (1) or
ionization (2) of the ions.  Both processes affect the lifetime of the stored
ions.  Both are pressure dependent. 
In the following, the associated rate coefficients for charge capture and
ionization are denoted as $\beta^1$ and $\beta^2$, respectively.

We can readily write out expressions for the time $t$ dependence of
the number of stored ions in the ring $N_{\rm{i}}$.  With the Cooler
off, this is given by
\begin{equation}
\frac{dN_{\rm i}^{\rm off}}{dt} = 
-N^{\rm off}_{\rm i} 
\left[(\beta^1 + \beta^2) \rho_{\rm c}^{\rm off} \eta_{\rm c} 
+ (\beta^1 + \beta^2) \rho_{\rm o}^{\rm off} (1-\eta_{\rm c}) 
+ \alpha^1_{\rm t} n_{\rm t} \eta_{\rm t} \right].
\label{eq:decoff}
\end{equation} 
Here $\eta_{\rm{c}}$ and $\eta_{\rm{t}}$ are the fractions of the ring
circumference covered by the Cooler and Target length, respectively,  $n_{\rm
t}$ is the Target electron density, and $\rho_{\rm c}$, $\rho_{\rm o}$
are the average residual gas densities in the Cooler (c) or in other sections
of TSR (o).  The ``off'' superscript is used to denote that the Cooler is off. 
With the Cooler on we have
\begin{equation}
\frac{dN_{\rm i}^{\rm on}}{dt} = 
-N^{\rm on}_{\rm i} 
\left[(\beta^1 + \beta^2) \rho_{\rm c}^{\rm on} \eta_{\rm c} 
+ (\beta^1 + \beta^2) \rho_{\rm o}^{\rm on} (1-\eta_{\rm c}) 
+ \alpha^1_{\rm t} n_{\rm t} \eta_{\rm t} 
+ \alpha^1_{\rm c} n_{\rm c} \eta_{\rm c} \right]
\label{eq:decon}
\end{equation} 
where the ``on'' superscript signifies the Cooler is on and $n_{\rm
  c}$ is the Cooler electron density.  The solution for these
equations is of the form
\begin{equation}
\label{eq:general}
N_{\rm i}^{\rm off/on}(t) = N_{\rm i,0}^{\rm off/on}\exp(-t/\tau^{\rm off/on})
\end{equation} 
where $N_{\rm i,0}^{\rm off/on}$ is the initial ion number and the ion
beam lifetimes $\tau^{\rm off/on}$ are given by
\begin{eqnarray}
(\tau^{\rm off})^{-1}  &=& 
(\beta^1 + \beta^2) \rho_{\rm c}^{\rm off} \eta_{\rm c} 
+ (\beta^1 + \beta^2) \rho_o^{\rm off} (1-\eta_{\rm c}) 
+ \alpha^1_{\rm t} n_{\rm t} \eta_{\rm t} 
\label{eq:tauoff}
\\
(\tau^{\rm on})^{-1} &=& 
(\beta^1 + \beta^2) \rho_{\rm c}^{\rm on} \eta_{\rm c} 
+ (\beta^1 + \beta^2) \rho_o^{on} (1-\eta_{\rm c})
+ \alpha^1_{\rm t} n_{\rm t} \eta_{\rm t} 
+ \alpha^1_{\rm c} n_{\rm c} \eta_{\rm c}.
\label{eq:tauon}
\end{eqnarray}

We can now readily solve for $\alpha_{\rm c}^1$ in terms of measured
quantities.  With the Cooler on and off, the count rates on
detectors 1 (recombination) and 2 (ionization) at a time $t$ are given
by
\begin{eqnarray}
R^{\rm 1,on}_{\rm c}(t) &=& N_{\rm i}^{\rm on}(t) (\eta_{\rm c} \alpha^1_{\rm c} n_{\rm c} + \beta^1 \rho_{\rm c}^{\rm on} \eta_{\rm c})
\label{le14}
\\
R^{\rm 2,on}_{\rm c}(t) &=& N_{\rm i}^{\rm on}(t) (\beta^2 \rho_{\rm c}^{\rm on} \eta_{\rm c})
\label{le15}
\\
R^{\rm 1,off}_{\rm c}(t) &=& N_{\rm i}^{\rm off}(t) (\beta^1 \rho_{\rm c}^{\rm off} \eta_{\rm c})
\label{le16}
\\
R^{\rm 2,off}_{\rm c}(t) &=& N_{\rm i}^{\rm off}(t) (\beta^2 \rho_{\rm c}^{\rm off} \eta_{\rm c}).
\label{le17}
\end{eqnarray} 
Combining Equations \ref{eq:tauoff} and \ref{eq:tauon} with Equations
\ref{le14}--\ref{le17} gives
 \begin{eqnarray}
  (\tau^{\rm on})^{-1} - (\tau^{\rm off})^{-1} &=& 
  \frac{R^{\rm 1,on}_{\rm c}(t)+ R^{\rm 2,on}_{\rm c}(t)}{N_{\rm i}^{\rm
on}(t)}
 - \frac{R^{\rm 1,off}_{\rm c}(t)+ R^{\rm 2,off}_{\rm c}(t)}{N_{\rm i}^{\rm
off}(t)}
\nonumber
\\
&&+ (\beta^1 + \beta^2)(\rho_{\rm o}^{\rm on}-\rho_{\rm o}^{\rm off}) (1-\eta_{\rm c}).
\label{le18}
\end{eqnarray} 
Direct pressure measurements do not show significant pressure changes in
TSR outside of the Cooler. Therefore we assume that the pressure in these
sections is independent of the state of the Cooler beam and thus $\rho_{\rm
o}^{\rm on} = \rho_{\rm o}^{\rm off}$.  We
take $t_0$ as the time when the Cooler is switched on or off which
gives $N_{\rm i}^{\rm on}(t_0) = N_{\rm i}^{\rm off}(t_0) \equiv N_{\rm
  i}(t_0)$.  Equation~\ref{le18} thereby simplifies to
\begin{equation}
  (\tau^{\rm on})^{-1} - (\tau^{\rm off})^{-1} = 
  \frac{R^{\rm 1,on}_{\rm c}(t_0)- R^{\rm 1,off}_{\rm c}(t_0) +
R^{\rm 2,on}_{\rm c}(t_0)- R^{\rm 2,off}_{\rm c}(t_0)}{N_{\rm i}(t_0)}.
 \label{le20}
\end{equation} 
Using Equations \ref{le14}--\ref{le17} to solve for $N_{\rm i}(t_0)$ gives
\begin{equation}
N_{\rm i}(t_0)  = \frac{R^{\rm 1,on}_{\rm c}(t_0) - R^{\rm 1,off}_{\rm
c}(t_0)\,R^{\rm 2,on}_{\rm c}(t_0) /  R^{\rm 2,off}_{\rm
c}(t_0) }{\eta_{\rm c} \alpha^1_{\rm c} n_{\rm c}}.
\label{le23}
\end{equation} 
Combining these last two equations we obtain
\begin{equation}
\hat{\alpha} \equiv \alpha^1_{\rm c} = \frac{(\tau^{\rm on})^{-1} - (\tau^{\rm
off})^{-1}}{\eta_{\rm c} n_{\rm c}}
   \frac{R^{\rm 1,on}_{\rm c}(t_0) - R^{\rm 1,off}_{\rm c}(t_0)\,R^{\rm
2,on}_{\rm c}(t_0) /  R^{\rm 2,off}_{\rm c}(t_0)} 
	{R^{\rm 1,on}_{\rm c}(t_0)- R^{\rm 1,off}_{\rm c}(t_0) + R^{\rm
2,on}_{\rm c}(t_0)- R^{\rm 2,off}_{\rm c}(t_0)}.
\label{eq:abs_adv}
\end{equation}

\begin{figure}
    \plotone{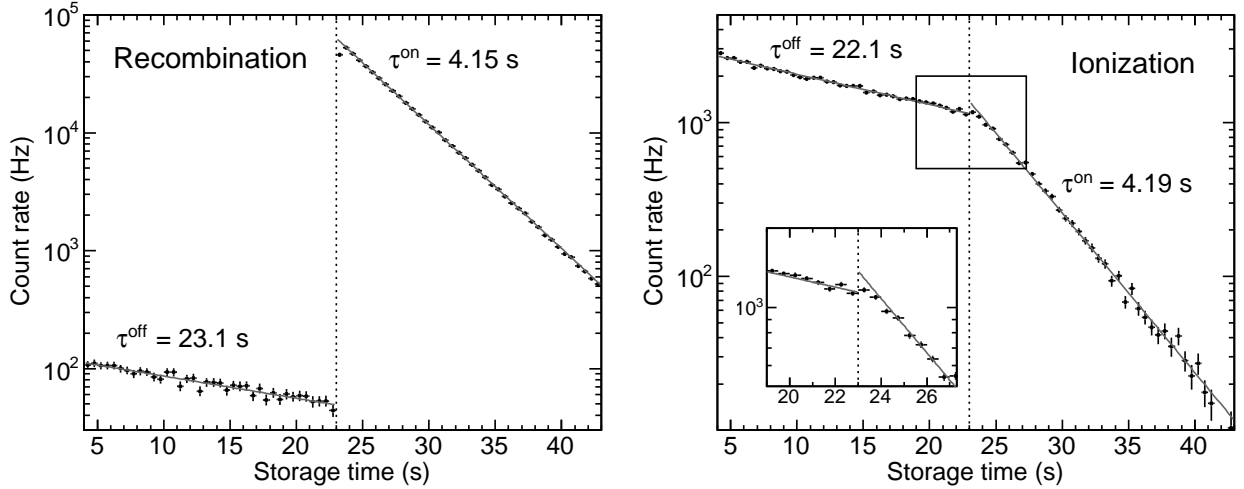}
    \caption{\label{lfAbs}
	Count rates versus storage time measured with the Cooler recombination
and ionization detectors, left and right, respectively.  The Cooler beam was
on for $t=0-3$~s, off for $3-23$~s, and switched back on at $t_0=23$~s. During
the on-phase, we matched the electron and ion velocities ($\hat{E}=0$).
 Data were not acquired during precooling ($t \le 3$~s).  The thin
solid lines indicate the exponential fits used to derive the ion beam lifetimes
$\tau^{\rm off}$ and $\tau^{\rm on}$.  The dotted vertical lines mark $t_0$. 
The inset in the right panel shows the increase of the ionization signal after
switching on the Cooler beam which causes an increase in the residual gas
pressure.
Extrapolations of the solid lines to $t_0$ were used to determine $R^{\rm
1,on}_{\rm c}$, $R^{\rm 2,on}_{\rm c}$, $R^{\rm 1,off}_{\rm c}$, and $R^{\rm
2,off}_{\rm c}$.  See Appendix~\ref{l:lifetime} for details.}
\end{figure}

The measured values used to solve Equation~\ref{eq:abs_adv} come from
data runs such as that shown in Figure~\ref{lfAbs}.  The lifetimes
$\tau^{\rm on}$ and $\tau^{\rm off}$ are obtained by fitting the
decaying recombination and ionization signals with the Cooler on and
off, respectively.  For a given state of the Cooler, the recombination
and ionization lifetimes agree to within their respective
uncertainties.  Here we use the lifetime measurement from the recombination data
as it has better statistics than that derived from the ionization data.  The
lifetime fits are extrapolated to $t_0$ in order to determine
$R^{\rm 1,on}_{\rm c}$, $R^{\rm 2,on}_{\rm c}$, $R^{\rm 1,off}_{\rm
  c}$, and $R^{\rm 2,off}_{\rm c}$.

The accuracy for the inferred value of $\hat{\alpha}$ as given by
Equation~\ref{eq:abs_adv} depends on the uncertainties in the various
measured quantities on the right hand side of the equation.  Here we
assume that for any variable $x$, the error $\sigma_x$ is uncorrelated
with other variables.  
We 
took partial derivatives to calculate $\sigma_{\hat{\alpha}}$ in a linear
approximation.
We simplify the notation using $R^{\rm
  1,on}_{\rm c}(t_0) \equiv R_{\rm 1,on}$, $\eta_{\rm c} \equiv \eta$, etc.
With the aid of MATHEMATICA, and after much algebraic manipulation, we find
\begin{eqnarray}
 \sigma^2_{\hat{\alpha}} & = & 
\left\{
\left[\eta_{}^2 n^2 R_{\rm 2,off}^2 (R_{\rm 1,off}-R_{\rm 1,on}+R_{\rm 2,off}-R_{\rm 2,on})^2 (R_{\rm 1,on} R_{\rm 2,off}-R_{\rm 1,off} R_{\rm 2,on})^2 (\sigma_{\tau_{\rm off}^{-1}}^2 + \sigma_{\tau_{\rm on}^{-1}}^2) \right.\right.\nonumber\\
& + &  R_{\rm 2,off}^2 (R_{\rm 1,off}-R_{\rm 1,on}+R_{\rm 2,off}-R_{\rm 2,on})^2 (R_{\rm 1,on} R_{\rm 2,off}-R_{\rm 1,off} R_{\rm 2,on})^2 (\sigma_{\eta_{}}^2 n^2 +\sigma_{n}^2\eta_{}^2)(\tau_{\rm off}^{-1}-\tau_{\rm on}^{-1})^2\nonumber\\
& + & \eta_{}^2 n^2 R_{\rm 2,off}^2 (R_{\rm 2,off}-R_{\rm 2,on})^2 (R_{\rm 1,on}+R_{\rm 2,on})^2 \sigma_{R_{\rm 1,off}}^2 (\tau_{\rm off}^{-1}-\tau_{\rm on}^{-1})^2\nonumber\\
& + & \eta_{}^2 n^2 R_{\rm 2,off}^2 (R_{\rm 1,off}+R_{\rm 2,off})^2 (R_{\rm 2,off}-R_{\rm 2,on})^2 \sigma_{R_{\rm 1,on}}^2 (\tau_{\rm off}^{-1}-\tau_{\rm on}^{-1})^2\nonumber\\
& + & \eta_{}^2 n^2 \left(R_{\rm 1,off} R_{\rm 2,on} (-R_{\rm 1,off}-2 R_{\rm 2,off}+R_{\rm 2,on})+R_{\rm 1,on} \left(R_{\rm 2,off}^2+R_{\rm 1,off} R_{\rm 2,on}\right)\right)^2 \sigma_{R_{\rm 2,off}}^2 (\tau_{\rm off}^{-1}-\tau_{\rm on}^{-1})^2\nonumber\\
& + & \left.\left. \eta_{}^2 n^2 (R_{\rm 1,off}-R_{\rm 1,on})^2 R_{\rm 2,off}^2 (R_{\rm 1,off}+R_{\rm 2,off})^2 \sigma_{R_{\rm 2,on}}^2 (\tau_{\rm off}^{-1}-\tau_{\rm on}^{-1})^2\right] \right.\nonumber\\
& / & \left.\left[\eta_{}^4 n^4 R_{\rm 2,off}^4 (R_{\rm 1,off}-R_{\rm 1,on}+R_{\rm 2,off}-R_{\rm 2,on})^4\right]\right\}.
\label{le28}
\end{eqnarray}
If detector 2 is not available, additional assumptions must be made,
a discussion of which is beyond the scope of this paper. 

To conclude we mention the special case described in
Section~\ref{l:abs} where the pressure in the Cooler does not change
with switching the electron beam ($\rho_{\rm c}^{\rm on} = \rho_{\rm
  c}^{\rm off}$) and the signal on detector 1 is dominated by electron
induced processes ($\alpha^1 \gg \beta^1$).  Equation~\ref{eq:abs_adv}
then reduces to
\begin{equation}
\hat{\alpha} = 
\frac{\tau_{\rm on}^{-1} - \tau_{\rm off}^{-1}}{\eta n}
\end{equation} 
where we have dropped the Cooler subscripts for convenience.  This is
equivalent to Equation~\ref{eqabs1}.  The associated error is given by
\begin{equation}
\sigma^2(\hat{\alpha}) = 
\frac{\sigma_{\tau_{\rm on}^{-1}}^2 + \sigma_{\tau_{\rm off}^{-1}}^2} {\eta^2 n^2} 
+ \frac{(n^2\sigma^2_{\eta} + \eta^2 \sigma^2_{n})(\tau_{\rm off}^{-1}-\tau_{\rm on}^{-1})^2}{\eta^4 n^4}.
\end{equation}

\section{List of abbreviations}
\begin{description}
\item[CP] collisionally ionized plasma
\item[DR] dielectronic recombination
\item[MCBP] multi-configuration Breit-Pauli
\item[MBRRC] merged-beams recombination rate coefficient
\item[PP] photoionized plasma
\item[PRRC] plasma rate coefficient
\item[RR] radiative recombination
\end{description}

\vfill
\eject

\bibliography{Fe11+_DR}

\begin{deluxetable}{lr}
\tablecaption{\label{table:fe11energylist}
	      Energy levels of \feelevenplus relative to the 
	      $3s^2\,3p^3\  [^4{S}^o_{3/2}]$ ground level
	      \citep{AtomicSpectraDatabase} for excitations within the 
	      M-shell ($\Delta \nc = 0$).
		}
\tablehead{\colhead{Level} & \colhead{Energy (eV)}}
\tablewidth{0pt}
\startdata
    $3s^2\,3p^3\ [^2{D}^o_{3/2}]$			& 5.1535\phn \\
    $3s^2\,3p^3\ [^2{D}^o_{5/2}]$			& 5.7126\phn \\

    $3s^2\,3p^3\ [^2{P}^o_{1/2}]$ 			& 9.1883\phn \\
    $3s^2\,3p^3\ [^2{P}^o_{5/2}]$ 			& 9.9826\phn \\

    $3s\,3p^4\ [^4{P}_{5/2}]$ 				& 34.0179\phn \\
    $3s\,3p^4\ [^4{P}_{3/2}]$ 				& 35.2121\phn \\
    $3s\,3p^4\ [^4{P}_{1/2}]$ 				& 35.7455\phn \\

    $3s\,3p^4\ [^2{D}_{3/2}]$ 				& 42.1571\phn \\
    $3s\,3p^4\ [^2{D}_{5/2}]$ 				& 42.3658\phn \\

    $3s\,3p^4\ [^2{P}_{3/2}]$ 				& 48.3174\phn \\
    $3s\,3p^4\ [^2{S}_{1/2}]$ 				& 48.8646\phn \\
    $3s^2\,3p^2\,(^1{D})\,3d\ [^2{P}_{3/2}]$ 		& 62.2153\phn \\
    $3s^2\,3p^2\,(^3{P})\,3d\ [^4{P}_{5/2}]$ 		& 63.5431\phn \\
    $3s\,3p^4\ [^2{P}_{1/2}]$ 				& 63.7093\phn \\

    $3s^2\,3p^2\,(^3{P})\,3d\ [^4{P}_{3/2}]$ 		& 64.0676\phn \\
    $3s^2\,3p^2\,(^3{P})\,3d\ [^4{P}_{1/2}]$ 		& 64.4433\phn \\

    $3s^2\,3p^2\,(^1{S})\,3d\ [^2{D}_{3/2}]$ 		& 65.2306\phn \\
    $3s^2\,3p^2\,(^1{S})\,3d\ [^2{D}_{5/2}]$ 		& 66.7085\phn \\

    $3s^2\,3p^2\,(^1{D})\,3d\ [^2{D}_{3/2}]$ 		& 68.6910\phn \\
    $3s^2\,3p^2\,(^1{D})\,3d\ [^2{D}_{5/2}]$ 		& 68.7629\phn \\
    $3s^2\,3p^2\,(^1{D})\,3d\ [^2{P}_{1/2}]$ 		& 70.5396\phn \\

    $3s^2\,3p^2\,(^3{P})\,3d\ [^2{F}_{5/2}]$ 		& 71.5066\phn \\

    $3s^2\,3p^2\,(^3{P})\,3d\ [^2{P}_{3/2}]$ 		& 71.6306\phn \\

    $3s^2\,3p^2\,(^1{D})\,3d\ [^2{S}_{1/2}]$ 		& 71.8650\phn \\
    $3s^2\,3p^2\,(^3{P})\,3d\ [^2{F}_{7/2}]$ 		& 72.0571\phn \\

    $3s^2\,3p^2\,(^3{P})\,3d\ [^2{D}_{5/2}]$ 		& 74.8778\phn \\
    $3s^2\,3p^2\,(^3{P})\,3d\ [^2{D}_{3/2}]$ 		& 75.0699\phn \\

\enddata
\end{deluxetable}

\clearpage

\begin{deluxetable}{crrr}
\tablewidth{0pt}
\tablecaption{\label{table:fe11integrals}Integrated DR rate coefficients for
\feelevenplus. Here, the values in brackets give the $1\sigma$ statistical
errors for the last digit(s) shown.}
\tablehead{\colhead{Energy range} & 
	   \colhead{$\int \alpha_{\rm theo}^{\rm DR}\, dE$} &
	   \colhead{$\int \alpha_{\rm exp}^{\rm DR}\, dE$} &
	   \colhead{$\kappa = \frac{\int \alpha_{\rm theo}^{\rm DR}\, dE}{\int
\alpha_{\rm exp}^{\rm DR}\, dE}$} \\
  	   \colhead{(eV)} & \multicolumn{2}{c}{($10^{-9}$ cm$^3$ s$^{-1}$ eV)}
}
\startdata	
$0.0135 - 0.45$	& 1.86 &	5.29(2)	& 0.352(6)\\
$0.45 - 5.5$	& 10.84 &	14.22(2)& 0.762(14) \\
$5.5 - 15.0$	& 3.41 &	4.61(1)	& 0.741(10) \\
$15.0 - 24.5$	& 2.81 &	3.70(2)	& 0.758(12) \\
$24.5 - 36.0$	& 4.20 &	5.11(1)	& 0.821(4) \\
$36.0 - 42.0$	& 1.20 &	1.89(1)	& 0.633(3) \\
$42.0 - 46.0$	& 0.98 &	1.57(1)	& 0.624(2) \\
$46.0 - 53.0$	& 2.32 &	2.59(1)	& 0.893(5) \\
$53.0 - 59.0$	& 3.62 &	2.58(1)	& 1.400(6) \\
$59.0 - 66.0$	& 14.04 &	8.30(1)	& 1.693(12) \\
$66.0 - 75.0$	& 0.21 &	0.44(1)	& 0.468(3) \\
$75.0 - 217.0$	& 3.51 &	5.16(3)	& 0.680(20) \\
$330.0 - 885.0$	& 1.89 &	6.7(2)	& 0.282(51)
\enddata
\end{deluxetable}

\clearpage

\begin{deluxetable}{ccc}
\tablecaption{Fit parameters $c_i$ (cm$^3$\,s$^{-1}$\,K$^{3/2}$)
		and $E_i$ (K) for the experimental DR PRRC  
		for \feelevenplus using 
		Equation~\ref{eqn:plasmafit_DR}.
		\label{table:plasmafit}
		}
\tablehead{ \colhead{$i$} & \colhead{$c_i$} & \colhead{$E_i$} }
\tablewidth{0pt}
\startdata
1	& $1.38\e{-3}$	& $9.48\e{2}$ \\
3	& $5.18\e{-3}$	& $5.61\e{3}$ \\
4	& $1.33\e{-2}$	& $1.92\e{4}$ \\
2	& $2.23\e{-2}$	& $6.14\e{4}$ \\
5	& $9.52\e{-2}$	& $2.70\e{5}$ \\
6	& $2.29\e{-1}$	& $8.28\e{5}$ \\
7	& $2.94\e{-1}$	& $4.90\e{6}$ \\ 
		
\enddata
\end{deluxetable}

%

\begin{figure}
    \plotone{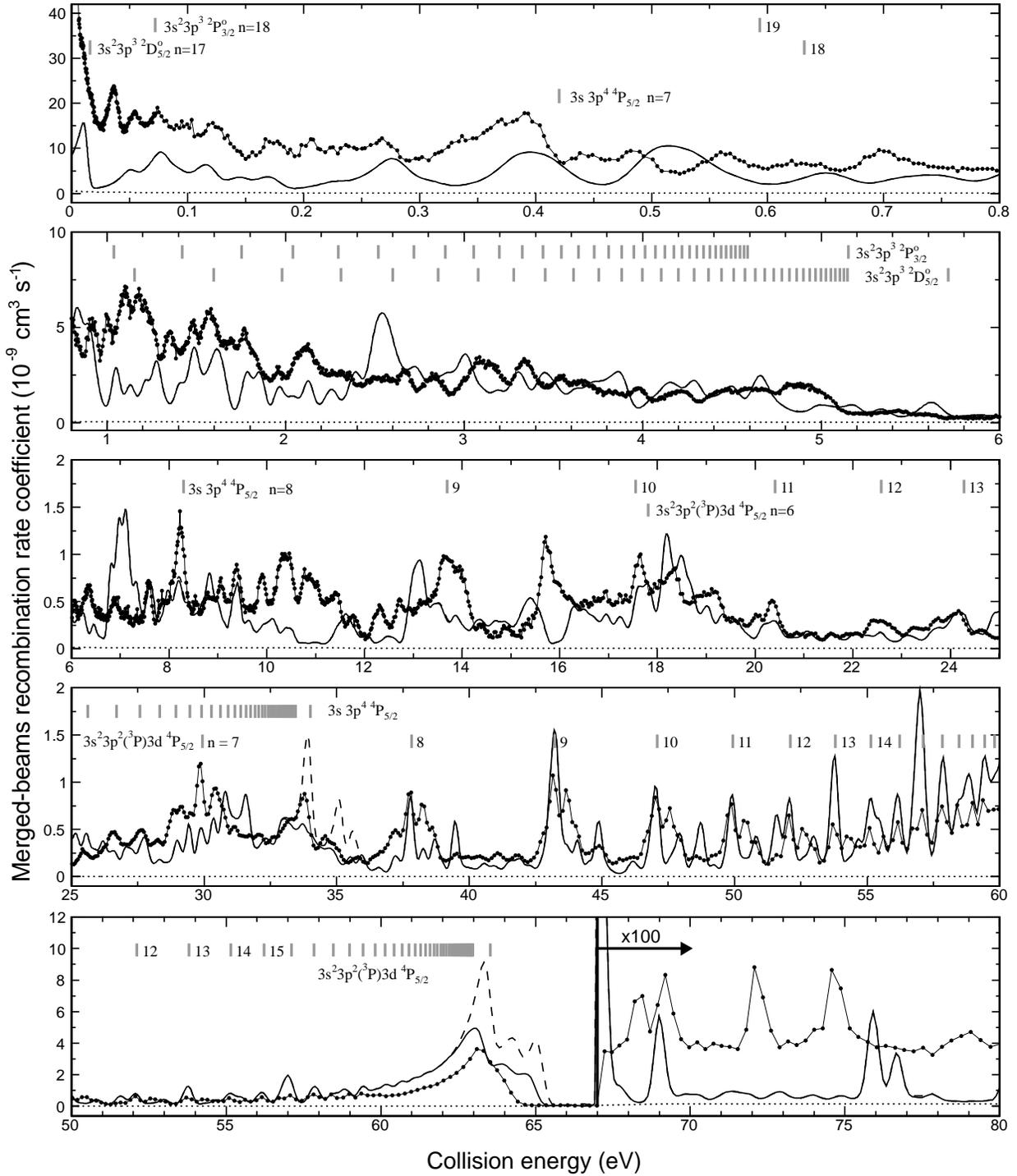}
    \caption{	\label{lfMBRC0}
	Caption on next page!
}
\end{figure}
\clearpage
\paragraph{Caption for Figure \ref{lfMBRC0}}
	MBRRC for \feelevenplus forming \fetenplus as a function of relative
collision energy. The data measured at the Cooler are shown by the connected
solid points.  The theoretical {\sc autostructure} results with field ionization
are shown by the solid line. Including the high $n$ contributions missing due to
field ionization gives the dashed line.  The theoretical RR MBRRC is shown by
the dotted line (on this scale it is almost compatible with zero at most
collision energies). For clarity, we show the DR resonance energies associated
with only four of the many possible Rydberg DR series (short vertical lines). We
label those series by the corresponding core excitation configuration. In each
series, the highest energy vertical mark corresponds to the series limit and the
penultimate mark to the approximate field ionization cut-off.
 
\begin{figure}
    \plotone{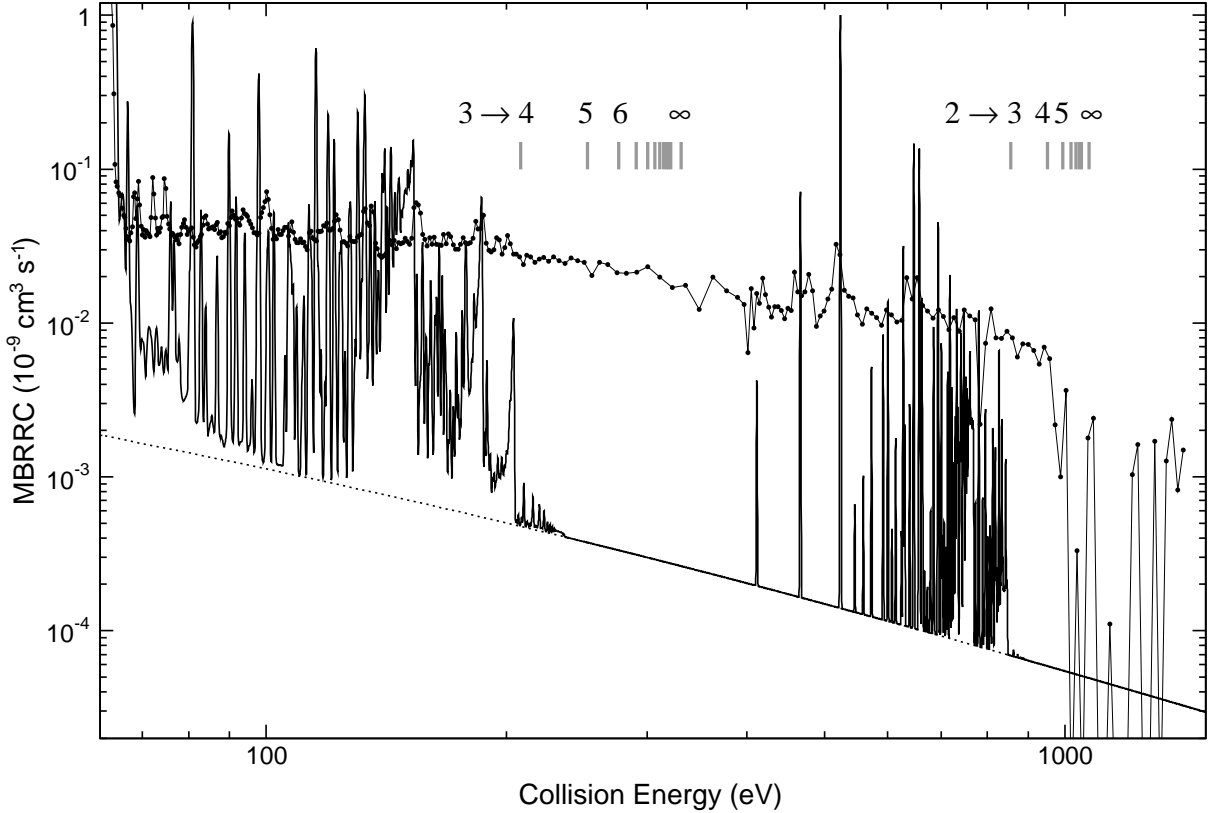}
    \caption{\label{lfMBRC1}
	Same as Figure \ref{lfMBRC0} but for collision energy ranges dominated
by $\Delta \nc>0$ transitions. The short vertical lines mark the DR series
limits for $3\rightarrow \nc'$ and $2\rightarrow \nc'$ core excitations as
calculated by using a hydrogenic approximation, assuming hydrogenic
Rydberg levels on a $3s^2\,3p^2$ core (labeled $3\rightarrow N'$) and on a
$2s^2\,2p^5\,3s^2\,3p^3$ core (labeled $2\rightarrow N'$).
No difference is seen between
the calculations with and without the field ionization effects included. A
significant amount of the measured DR flux is due to channels not accounted for
in the theoretical calculations.
}
\end{figure}

\begin{figure}
    \plotone{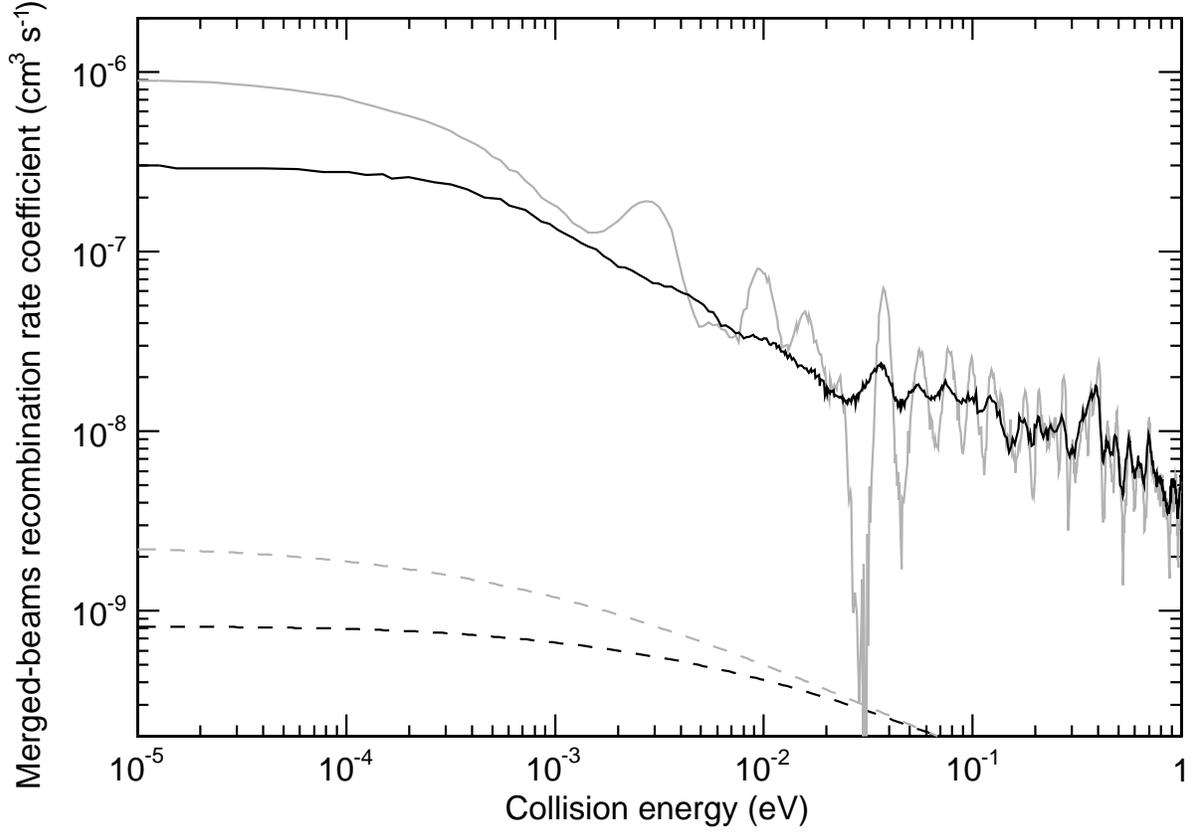}
    \caption{\label{lfMBRC3}
	Comparison of the low energy \feelevenplus to \fetenplus MBRRC data
acquired at the Cooler (black full line, $\ktperp^{\rm{c}}\approx13.5$~meV) and
at the Target (gray full line, $\ktperp^{\rm{t}}\approx1.5$~meV). 
The dashed lines show the theoretical RR contribution convolved with the
electron energy spreads of the Cooler (black) and Target (gray), respectively. }
\end{figure}

\begin{figure}
    \plotone{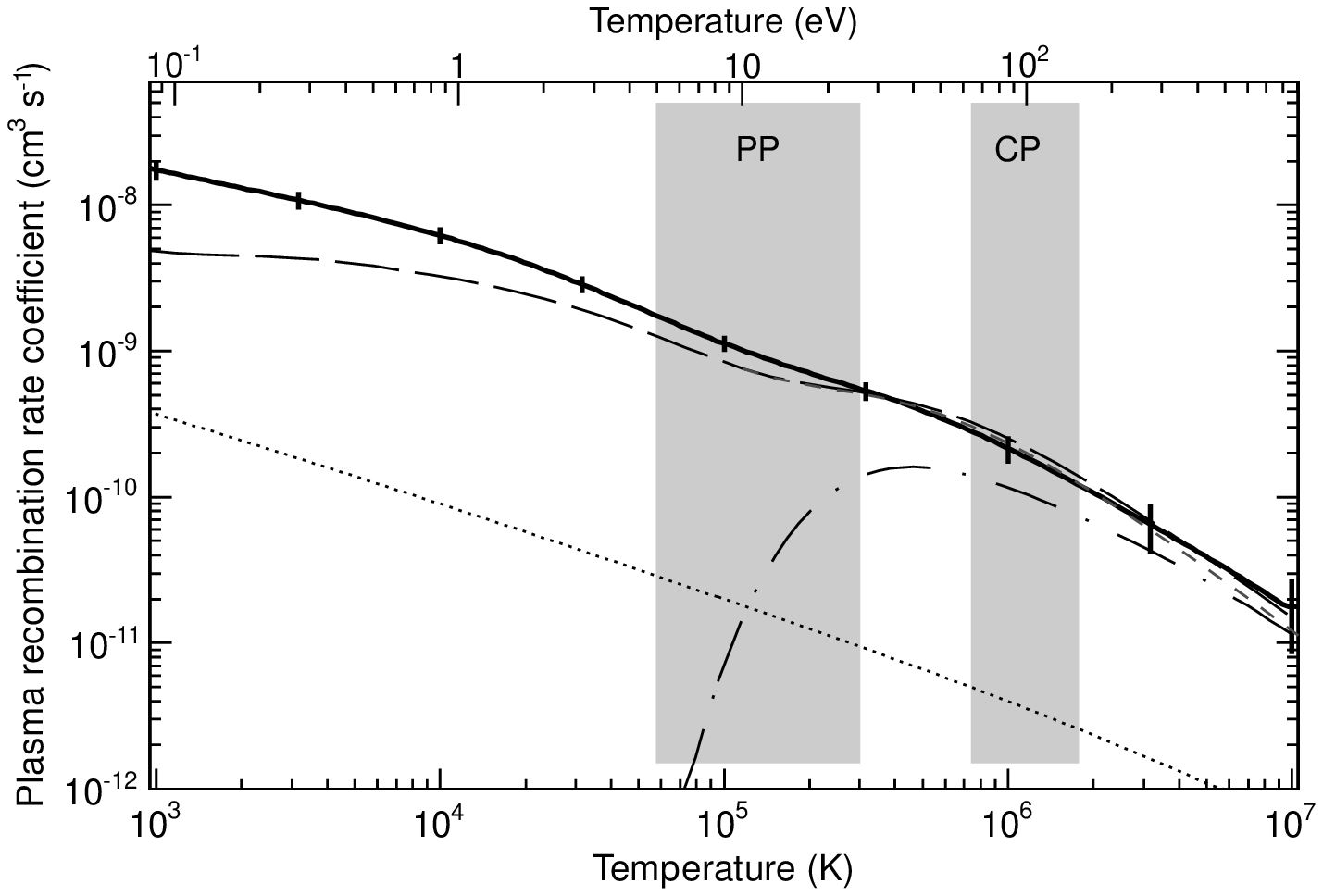}
    \caption{\label{fig:fe11_plasma}
	Comparison of the experimental and theoretical DR PRRC for \feelevenplus
forming \fetenplus.
    	The thick solid line gives the experimental results and the error bars 
        the experimental uncertainty at a $1\sigma$ confidence 
        level.  The previously recommended rate coefficient of
\cite{ArnaudRaymond1992}
	is shown by the long-dash-dotted curve.  
	The short dashed curve gives previous results of \cite{Badnell2006b} 
	while the long dashed curve presents our new results 
	which extend these older calculations by including $\nc = 3 \to 4$ and
$2 \to 3$ core excitations.
	These two curves overlap below $\sim 3 \times 10^5$~K. For comparison we
plot also the calculated RR PRRC (dotted line) 
of \cite{Badnell2006b}.
    	The shaded areas indicate the plasma temperatures where the 
	\feelevenplus abundance is $\ge 1\%$ in photoionized plasmas
	\citep[PP;][]{Kallman2010} and in collisionally ionized plasmas
	\citep[CP;][]{Bryans2006, Bryans:ApJ:2009}. 
    }
\end{figure}

\end{document}